\documentstyle[epsf,aps]{revtex}  
\begin{document}
\draft         %

\preprint{DFNAE-IF/UERJ-98/05}

\title{Production of a Single and a Pair of Heavy-Exotic-Leptons Through the pp Collisions}

\author{ J.\ E.\ Cieza Montalvo $^*$}

\author{P. P. de Queiroz Filho $^\dagger$}

\address{Instituto de F\'{\i}sica, Universidade do Estado do Rio de Janeiro 
CEP $20559-900$ Rio de Janeiro, Brazil}


\maketitle


\begin{abstract}
We investigate the production and signatures of a single, pair of heavy-exotic leptons via the electron-positron induced process and a pair of heavy exotic leptons via the Drell-Yan process at the CERN Large Hadron Collider (LHC). We show that the cross-section for the case of the production of heavy pair exotic leptons are competitive with the ones for the single production of exotic leptons. We also exhibit some kinematical distributions.


\vskip 2.5cm

{\em Submitted to Phys.\ Rev.\ D}

\end{abstract}

\newpage



\section{INTRODUCTION}

The standard electroweak theory (SM) provides a very satisfactory description of most elementary particle phenomena up to the presently available energies. However, the SM has many problems: the spectrum of elementary fermions, no unifications of the forces, violation of CP and the existence of many parameters. Concerning to the first problem, there is no theoretical explanation for the existence of several generations and for the values of the masses.  It was established at the CERN $e^{+} e^{-}$ collider LEP that the number of light neutrinos is three \cite{lep}.  However this result does not impede the assumption that further generations coluld exist \cite{nov}.

Many models, such as composite models \cite{af,bu1}, grand unified theories \cite{la}, technicolor models \cite{di}, superstring-inspired models \cite{e6}, mirror fermions \cite{maa}, predict the existence of new particles with masses around of the scale of $1$ TeV and they consider the possible existence of new generations of fermions.

In this work, we shall consider the possibility of existence of a new generation of leptons and analyze the production of exotic heavy leptons in the large hadron collider LHC ($\sqrt{s} = 14$ TeV). We present here another form to produce these leptons, namely through the reactions $pp \rightarrow \gamma \gamma \rightarrow e^{-} e^{+} \rightarrow  L^{-} L^{+} (L^{-} e^{+}, L^{0} \nu)$, and we show that the production of a pair of exotic leptons are competitive with the ones for the single heavy lepton, that is so, due to the vector component of the Z-boson coupling to exotic leptons and to the exchange of a photon in the $s$ channel, although the phase space is more reduced for pair production than for single heavy lepton production.

We also study the production of pair of exotic leptons through the Drell-Yan mechanism and we will see that this production is dominant over the single heavy lepton production \cite{sim3}.

The three models that we consider here include new fermionic degrees of freedom, which introduce naturally a number of unknown mixing angles and fermionic masses \cite{tom}. These models are: the vector singlet model (VSM) \cite{gon}, vector doublet model (VDM) \cite{riz} and the fermion-mirror-fermion (FMF) model \cite{maa}.

Exotic leptons mixed with the standard leptons interact through the
standard weak vector bosons $W^{+}, W^{-}$ and $Z^{0}$, according to the interacting Lagrangians

\begin{equation}
{\cal L}_{\rm NC} = \frac{g}{4 cos\theta_{W}} \left[{ \bar{F}_{i}}
\gamma^{\mu} (g_{V}^{ij} - g_{A}^{ij} \gamma^{5}) F_{j} +
{\bar{F}_{i}} \gamma^{\mu} (g_{V}^{ij} - g_{A}^{ij} \gamma^{5}) f_{j} \right]
Z_{\mu} 
\label{lag1}
\end{equation}
   
and 

\begin{equation}
{\cal L}_{\rm CC} = \frac{g}{2 \sqrt{2}} { \bar{L}^{0} \gamma^{\mu} 
(C_{V}^{ij} - C_{A}^{ij} \gamma^{5}) e^{-} W_{\mu} } ,
\label{lag2}
\end{equation}
where $g_{V}^{ij}$ and $g_{A}^{ij}$ are the corresponding neutral
vector-axial coupling constants and $C_{V}^{ij}$ and $C_{A}^{ij}$ are
the charged vector-axial coupling constants, which are given in \cite{cie,sim1} for each of the three models that we study here.

In this work we deal only with the lepton sector. In \cite{cie,sim2} it is shown that for the process $e^{-}e^{+} \rightarrow L^{-} L^{+}$ is a difficult task to distinguish the VSM and VDM leptons since they have similar cross sections and angular distributions. This is not the case for FMF leptons, whose cross section and angular distribution  is different from the other class of leptons. In this work we will show that, for the process $pp \rightarrow qq \rightarrow L^{-} L^{+}$ will be possible to distinguish the VSM leptons from the VDM and FMF leptons.

We consider here that all mixing angles have the same value $\theta_{i}
= 0.1$, although phenomenological analysis \cite{tom1} gives an upper
bound of $sin^{2} \theta_{i} \leq 0.03$. This means that the value of
$\theta_{i}$ can be scaled up to $0.173$.

The outline of this paper is the following. The luminosities of $e^{-} e^{+}$ for $pp$ colliders are given in Sec. II. In Sec. III we study the production of a single and a pair of exotic-lepton. We summarize the results in Sec. IV.


\section{Luminosities for Hadronic Collisions in the Elastic Process}    

We analyze as the case of inelastic scattering as the elastic one. For the elastic scattering, where the protons remain intact the calculation gives 

\begin{eqnarray}
\frac{d \rm L^{el}}{d\tau}  =  \int_{\tau}^{1} \frac{dx_{1}}{x_{1}} \int_{x_{1}}^{1} \frac{dx_{3}}{x_{3}} f_{\ell/\gamma} (x_{3})    
f_{\gamma/p} (x_{1}/x_{3}) \int_{x_{2}}^{1} \frac{dx_{5}}{x_{5}} 
f_{\ell/\gamma} (x_{5}) f_{\gamma/p} (x_{2}/x_{5})  \; ,
\end{eqnarray}  
where  $\tau = x_{1} x_{2}$.

Here $f_{\ell/\gamma} (x)$ is the effective photon approximation for the lepton into the photon and $f_{\gamma/p} (x)$ is the equivalent-photon distribution for elastic $pp$ scattering and we take \cite{kni} 

\begin{eqnarray}
f_{\gamma/p} (x) = &&\frac{\alpha}{2 \pi} x \Bigl [c_{1} y \ln(1+ \frac{c_{2}}{z})- (y+ c_{3}) \ln(1- \frac{1}{z})+ \nonumber  \\
&&\frac{c_{4}}{z- 1}+ \frac{c_{5} y+ c_{6}}{z}+ 
\frac{c_{7}y+ c_{8}}{z^{2}}+ \frac{c_{9}y+ c_{10}}{z^{3}} \Bigr ] \; ,
\end{eqnarray} 
where y and z depend on x,

\[
y = \frac{1}{2}- \frac{2}{x^2}, z = 1+ \frac{a}{4} \frac{x^{2}}{1- x}
\]
and $c_{i} (i = 1..  10)$ are constants.

In Fig. $1$, we present numerical examples for the differential $e^{-} e^{+}$ luminosities for the case of elastic $pp$ scattering at the LHC $[ \sqrt{s} = 14 TeV ]$.


\section{CROSS SECTION PRODUCTION}

We begin our study with the mechanism of the Drell-Yan production of pair of exotic leptons, that is we study the process $pp \rightarrow q   \bar{q} \rightarrow L^{-} L^{+}$. This process take place through the exchange of a boson $Z^{0}$ and $\gamma$ in the $s$ channel. 

Using the interaction Lagrangians, Eqs. ($1$) and ($2$), we evaluate the cross section for the process $q \bar{q} \rightarrow L^{-} L^{+}$ obtaining:
  
\begin{eqnarray} 
\left (\frac{d \sigma}{d\cos \theta} \right )_{L^{+} L^{-}} = &&\frac{\beta \alpha^{2} \pi}{s^{2}} \Biggl [\frac{e_{q}^{2}}{s} ( 2 s M_{L}^{2} + (M_{L}^{2} - t)^{2} + (M_{L}^{2} - u)^{2}  )  \nonumber \\
&&+ \frac{e_{q}}{2 \sin^{2} \theta_{W} \cos^{2} \theta_{W} (s - M_{Z}^{2})} ( 2 s M_{L}^{2} g_{V}^{LL} g_{V}^{l}  \nonumber \\
&&+g_{V}^{LL} g_{V}^{l} ( (M_{L}^{2} - t)^{2} + (M_{L}^{2} - u)^{2}  )  + g_{A}^{LL} g_{A}^{l} ( (M_{L}^{2} - u)^{2} - (M_{L}^{2} - t)^{2} )  \Biggr ] \nonumber \\
&&+ \frac{\beta \pi \alpha^{2}}{16 s \cos^{4} \theta_{W} \sin^{4} \theta_{W}} \frac{1}{(s - M_{Z}^{2} )^{2}} \Biggl [(g_{V}^{{LL}^{2}} + g_{A}^{{LL}^{2}}) (g_{V}^{{l}^{2}} + g_{A}^{{l}^{2}}) ((M_{L}^{2} - u)^{2}  \nonumber \\
&&+(M_{L}^{2} - t)^{2}) + 2 (g_{V}^{{LL}^{2}} - g_{A}^{{LL}^{2}}) (g_{V}^{{l}^{2}} + g_{A}^{{l}^{2}}) s M_{L}^{2}   \nonumber \\
&&+ 4 g_{V}^{LL} g_{A}^{LL} g_{V}^{l} g_{A}^{l} ((M_{L}^{2} - u)^{2} - (M_{L}^{2} - t)^{2}) \bigr )   \Biggr] \; ,
\end{eqnarray}
where $\beta_{L} = \sqrt{1- 4 M_{L}^{2}/s}$ is the velocity of
exotic-fermion in the c.m. of the process, $M_{Z}$ is the mass of the $Z$ boson, $\sqrt{s}$ is
the center of mass energy of the $q \bar{q}$ system, $t = M_{L}^{2}
- \frac{s}{2} (1 - \beta \cos \theta)$ and {} $u = M_{L}^{2} - \frac{s}{2}
(1 + \beta \cos \theta)$, where $\theta$ is the angle between the exotic
lepton and the incident quark, in the c.m. frame.

The total cross section for the process $pp \rightarrow qq \rightarrow 
L^{-} L^{+}$ is related to the subprocess $qq \rightarrow L^{-} L^{+}$ 
total cross section $\hat{\sigma}$, through

\begin{equation}  
\sigma =
\int_{\tau_{min}}^{1}
\int_{\ln{\sqrt{\tau_{min}}}}^{-\ln{\sqrt{\tau_{min}}}} d\tau dy 
q(\sqrt{\tau}e^y, Q^2) q(\sqrt{\tau}e^{-y}, Q^2)  \hat{\sigma}(\tau, s)  \; , 
\end{equation}
where $\tau = \frac{\hat{s}}{s} (\tau_{min} = \frac{4 M_L^2}{s})$,
with $s$ being the center-of mass  energy  of the $pp$ system and $q(x,Q^2)$ is the quark structure function.

Another form to produce a single or a pair of heavy exotic leptons is via the electron-positron induced process, namely through the  reaction of the type $pp \rightarrow \gamma \gamma \rightarrow e^{-} e^{+} \rightarrow L^{-} L^{+}  (L^{-} e^{+}, L^{0} \nu) $.

We first start with the pair production of exotic leptons. We study the subprocess   $e^{-} e^{+} \rightarrow L^{-} L^{+} $, provided that there is enough available energy ($\sqrt{s} \geq 2M_{L}$). This subprocess take place through the exchange of a photon in the $s$ channel, boson $Z^{0}$ in the $s$ and $t$ channel and a boson $W$ in the $t$ channel. Using the interaction Lagrangians of Sec. II, the cross section for the subprocess $e^{-} e^{+} \rightarrow L^{-} L^{+}$, involving a neutral current was already evaluated in \cite{cie}.

The production of a single charged heavy exotic lepton can be studied through the neutral current reaction $e^{-} e^{+} \rightarrow L^{-} e^{+}$, involving a boson $Z$ in the $s$ and $t$ channel, then using  the interactions Lagrangians, Eqs. ($1$) and ($2$), we obtain:

\hskip 0.5cm

\begin{eqnarray} 
\left (\frac{d \sigma}{d\cos \theta} \right )_{L^{-} e^{+}} = &&\frac{(1- M_{L}^{2}/s) G_{f}^{2} M_{Z}^{4}}{8 \pi s}  \Bigl [ \frac{1}{(s - M_{Z}^{2})^{2}}  ({g_{V}^{Le}}^{2} + {g_{A}^{Le}}^{2}) ({g_{V}^{l}}^{2} + {g_{A}^{l}}^{2})  \nonumber \\
&&\bigl (-u (M_{L}^{2} - u) - t (M_{L}^{2} - t) \bigr ) + 4 g_{V}^{Le} g_{A}^{Le} g_{V}^{l} g_{A}^{l}  \bigl (- u (M_{L}^{2} - u) + t (M_{L}^{2} - t) \bigr ) \nonumber \\  
&&+\frac{1}{(t - M_{Z}^{2})^{2}} ({g_{V}^{Le}}^{2} + {g_{A}^{Le}}^{2}) ({g_{V}^{l}}^{2} + {g_{A}^{l}}^{2}) \bigl ( s(s - M_{L}^{2}) - u(M_{L}^{2} - u) \bigr )  \nonumber \\
&&+ 4 g_{V}^{Le} g_{A}^{Le} g_{V}^{l} g_{A}^{l} \bigl (-u(M_{L}^{2} - u) - s(s- M_{L}^{2}) \bigr )  \nonumber  \\
&&+\frac{2}{(s- M_{Z}^{2})} \frac{1}{(t- M_{Z}^{2})} u (M_{L}^{2} - u) \bigl ( ({g_{V}^{Le}}^{2} + {g_{A}^{Le}}^{2})  \nonumber \\
&&({g_{V}^{l}}^{2} + {g_{A}^{l}}^{2}) + 4g_{V}^{Le} g_{A}^{Le} g_{V}^{l} g_{A}^{l} ) \bigr )  \Bigr] \; , 
\end{eqnarray}
where $\beta_{L} = \frac{1- M_{L}^{2}/s}{1+ M_{L}^{2}/s}$ is the velocity of exotic-fermion in the c.m. of the process, $G_{F}$ is the Fermi
coupling constant, $M_{Z}$ is the mass of the $Z$ boson, $\sqrt{s}$ is
the center of mass energy of the $e^{+} e^{-}$ system, $t = M_{L}^{2} - \frac{s}{2} {(1 + \frac{M_{L}^{2}}{s}) - (1 - \frac{M_{L}^{2}}{s})  \cos\theta}$ and {} $u = M_{L}^{2} - \frac{s}{2} {(1 + \frac{M_{L}^{2}}{s}) + (1 - \frac{M_{L}^{2}}{s}) \cos \theta}$, where $\theta$ is the angle between the exotic lepton and the incident electron, in the c.m. frame.  This cross section for the subprocess $e^{-} e^{+} \rightarrow L^{-} e^{+}$ is equal to the one calculated in \cite{sim1}.

Now, the production of exotic neutrino can be studied through the analysis of the charged current reaction of the type $e^{-} e^{+} \rightarrow L^{0} \nu$, involving the  exchange of a boson $W^{\pm}$ in the $t$ channel. Once more, using the interaction Lagrangians, Eqs. ($1$) and ($2$),  we evaluate the cross section for the process $e^{-} e^{+} \rightarrow L^{0} \nu$, from which we obtain:

\vskip 0.5cm

\begin{eqnarray} 
\left (\frac{d \sigma}{d\cos \theta} \right )_{L^{0} \nu} = &&\frac{(1- M_{L}^{2}/s) G_{f}^{2} M_{W}^{4}}{32 \pi s (t - M_{W}^{2})^{2}}  \Bigl [ s(s- M_{L}^{2}) \bigl ( ({g_{V}^{L^{0} \nu}}^{2} + {g_{A}^{L^{0} \nu}}^{2}) ({g_{V}^{l}}^{2} + {g_{A}^{l}}^{2}) \nonumber \\
&&- 4 g_{V}^{L^{0} \nu} g_{A}^{L^{0} \nu} g_{V}^{l} g_{A}^{l} \bigr )
- u(M_{L}^{2} - u) \bigl ( ({g_{V}^{L^{0} \nu}}^{2} + {g_{A}^{L^{0} \nu}}^{2}) ({g_{V}^{l}}^{2} + {g_{A}^{l}}^{2}) \nonumber \\
&&+ 4 g_{V}^{L^{0} \nu} g_{A}^{L^{0} \nu} g_{V}^{l} g_{A}^{l} \bigr ) \Bigr ] \; ,
\end{eqnarray}
where $M_{W}$ is the mass of the $W$ boson and $L^{0}$ is the neutrino. We note that this cross section for the subprocess $e^{-} e^{+} \rightarrow L^{0} \nu$ is equal to the one evaluated in \cite{sim4}.

In other to obtain the total cross section for the production of a pair of heavy exotic leptons ($pp \rightarrow L^{-} L^{+} (L^{0} L^{0})$), or a single  heavy exotic lepton ($pp \rightarrow L^{-} e^{+} (L^{0} \nu)$), we must fold the $\hat{\sigma}_{e^{-} e^{+} \rightarrow L^{-} L^{+} (L^{-} e^{+}, L^{0} \nu)}$ with the $\gamma$ ($f_{\gamma/p}$) and $e^{\pm}$ ($f_{e/\gamma}$) distributions in the proton: 

\begin{eqnarray} 
\sigma^{el} (s) =&&\int_{\tau_{min}}^{1} d\tau \frac{d \rm L^{el}}{d\tau} \hat{\sigma} (\hat{s} = \tau s ) \;,
\end{eqnarray} 
where $\frac{d \rm L^{el}}{d\tau}$ is the differential luminosity given above and $\hat{s} = \tau s$ (s) being the center-of-mass energy squared of the subprocess (process).

\hskip 0.5cm


\section{RESULTS AND CONCLUSIONS}

The process for the heavy lepton production in hadronic colliders was well studied in the literature \cite{dic} and was shown that the dominant contribution are the well known Drell-Yan process and gluon-gluon fusion \cite{scot,dic}.

We present in Fig. $2$ the cross section for the process 
$pp \rightarrow qq \rightarrow L^{-} L^{+}$, involving the three models considered here: VSM,  VDM and FMF models, for the LHC ($14$ TeV), using the mechanism of Drell-Yan. In all calculations we take $\sin^{2}_{\theta_W} = 0.2315$, $M_Z = 91.188$ and $M_{W} = 80.33$. Considering that the expected integrated luminosity for the LHC will be of order of $\propto 10^{5} pb^{-1}/yr$ and taking the mass of the  lepton  equal to $200$ GeV, we have a total of $ \simeq 175.10^{2}$ leptons pairs produced per year, for the VSM, $548.10^{2}$ and $684.10^{2}$ for the VDM and for the FMF leptons. It is seem not a difficult matter to distinguish the VSM leptons from the other two class of leptons. However this is not the case between VDM leptons and FMF leptons, because the difference between these two classes of particles are only $\simeq 25 \%$.

Fig. $3$ shows the angular distribution of heavy lepton of mass equal to $500$ GeV at the LHC for the process $pp \rightarrow qq \rightarrow L^{-} L^{+}$. We can observe from this result that this distribution is symmetric for both the VSM and VDM and it gives a peaked distribution for $\cos \theta \simeq -1$ for the FMF model.

Also, as seen from fig. $2$, through the cross sections, it will be possible to distinguish the VSM leptons from the other types of leptons. On the other hand, from fig. $3$, it will be  possible the distinguishion the VDM leptons from the FMF leptons through the distribution cross section. This way we will be able to distinguish the  three types of leptons, the VSM the VDM and the FMF. We then have a well defined signals for the VSM and VDM leptons, remembering that for the case $e^{-} e^{+} \rightarrow L^{-} L^{+}$ it  was not possible to differentiate VSM leptons from VDM leptons \cite{cie}. We still note here that this production mechanism, via the Drell-Yan, for exotic leptons are larger than the production of single exotic leptons \cite{sim3}.

In Fig. $4$ we show the cross section for the pair production of exotic  leptons $pp \rightarrow \gamma \gamma \rightarrow e^{-} e^{+} \rightarrow L^{-} L^{+}$ of mass equal to $200$ GeV for the VSM, VDM and FMF models at the LHC. We see from these results that we can expect for each of these models a total of around $\simeq  20$ heavy leptons pairs produced per year.  These productions mechanism are not competitive with the Drell-Yan productions.

Fig. $5$ shows the angular distribution of heavy lepton of mass equal to $500$ GeV at the LHC for the process $pp \rightarrow \gamma \gamma \rightarrow e^{-} e^{+} \rightarrow L^{-} L^{+}$. We can observe from this result that this distribution is fairly equal for the three types of leptons: VSM, VDM and FMF.

Fig. $6$ shows the cross section for the process $pp \rightarrow \gamma \gamma \rightarrow e^{-} e^{+} \rightarrow L^{-} e^{+}$, for the LHC. We note that for the case of FMF leptons we will have a total of $20$ exotic leptons produced per year, while for the other cases, that is, for the VSM and VDM models, we will have only $5$ exotic leptons produced per year. To obtain these results we have take a mass of the exotic lepton equal to $200$ GeV, for the three models: VSM, VDM and FMF, respectively. We note here that for the case of Drell-Yan production which was evaluated in \cite{sim3}, gives a total of $\propto 10^{4}$ exotic leptons produced per year.

Fig. $7$ shows the angular distribution of heavy lepton of mass equal to $500$ GeV at the LHC for the process $pp \rightarrow \gamma \gamma \rightarrow e^{-} e^{+} \rightarrow L^{-} e^{+}$. We can observe from this result that it gives a peaked distribution at $\cos \theta  \simeq + 1$ for the three models: VSM, VDM and FMF. There we expect the $L^{-}$ to go predominantly in the direction of the original proton for all three models.

Fig. $8$ shows the cross section for the production of exotic neutrinos, at the LHC, through the reaction $pp \rightarrow \gamma \gamma \rightarrow e^{-} e^{+}   \rightarrow L^{0} \nu$. We observe from these results that we have for the VSM a total of around $17$ exotic neutrinos produced per year, while for the VDM and FMF models we have a total of $34$ and $57$, respectively, exotic neutrinos produced per year. These results are obtained for a lepton mass equal to $200$ GeV.

Fig. $9$ shows the angular distribution of heavy neutrino of mass equal to $500$ GeV at the LHC for the process $pp \rightarrow \gamma \gamma \rightarrow e^{-} e^{+} \rightarrow L^{0} \nu$. We can observe from this result that it gives a peaked distribution for $\cos \theta  \simeq + 1$, for the three models: VSM, VDM and FMF, and it is  more open for $\cos \theta \simeq -1$, that is, the $L^{-}$ go in the oposite direction to the one of the original proton.

\section{Acknowledgments}
I would like to thank Prof. R. O. Ramos for a careful reading of the manuscript.
\newpage




\newpage

\begin{center}
FIGURE CAPTIONS
\end{center}

\vspace{0.5cm}

{\bf Figure 1}: Differential $e^{-} e^{+}$ luminosities for the case of  elastic $pp$ scattering at the LHC $[ \sqrt{s} = 14]$ TeV.

{\bf Figure 2}: Total cross section for the process $pp  \rightarrow qq \rightarrow  L^{+} L^{-}$ as a function of $M_{L}$ at $s = 14$ TeV: 
(a) vector singlet model (dotted line); (b) vector doublet model
(dashed line); (c) fermion mirror fermion (solid line).

{\bf Figure 3}: Angular distribution of the exotic lepton of mass equal to $500$ GeV at the LHC, for the process $pp  \rightarrow qq \rightarrow  L^{+} L^{-}$  for the (a) vector singlet model (dotted line); (b) vector doublet model (dashed line); (c) fermion mirror fermion (solid line).

{\bf Figure 4}: Total cross section for the process $pp \rightarrow \gamma \gamma  \rightarrow e^{-} e^{+} \rightarrow L^{-} L^{+}$ as a function of $M_{L}$ at $s = 14$ TeV: (a) vector singlet model (dotted line); (b) vector doublet model (dashed line); (c) fermion mirror fermion (solid line).

{\bf Figure 5}: Angular distribution of the exotic lepton of mass equal to $500$ GeV at the LHC, for the process $pp  \rightarrow \gamma \gamma \rightarrow e^{-} e^{+} \rightarrow L^{-} L^{+}$  for the (a) vector singlet model (dotted line); (b) vector doublet model (dashed line); (c) fermion mirror fermion (solid line).

{\bf Figure 6}: Total cross section for the process $pp \rightarrow \gamma \gamma  \rightarrow e^{-} e^{+} \rightarrow L^{-} e^{+}$ as a function of $M_{L}$ at $s = 14$ TeV: (a) vector singlet model (dotted line); (b) vector doublet model (dashed line); (c) fermion mirror fermion (solid line).

{\bf Figure 7}: Angular distribution of the exotic lepton of mass equal to $500$ GeV at the LHC, for the process $pp  \rightarrow \gamma \gamma \rightarrow e^{-} e^{+} \rightarrow L^{-} e^{+}$  for the (a) vector singlet model (dotted line); (b) vector doublet model (dashed line); (c) fermion mirror fermion (solid line).

{\bf Figure 8}: Total cross section for the process $pp \rightarrow \gamma \gamma  \rightarrow e^{-} e^{+} \rightarrow L^{0} \nu$ as a function of $M_{L}$ at $s = 14$ TeV: (a) vector singlet model (dotted line); (b) vector doublet model (dashed line); (c) fermion mirror fermion (solid line).

{\bf Figure 9}: Angular distribution of the exotic lepton of mass equal to $500$ GeV at the LHC, for the process $pp  \rightarrow \gamma \gamma \rightarrow e^{-} e^{+} \rightarrow L^{0} \nu$  for the (a) vector singlet model (dotted line); (b) vector doublet model (dashed line); (c) fermion mirror fermion (solid line).

\newpage

\begin{figure}[b]
\epsfysize=18cm 
{\centerline{\epsfbox{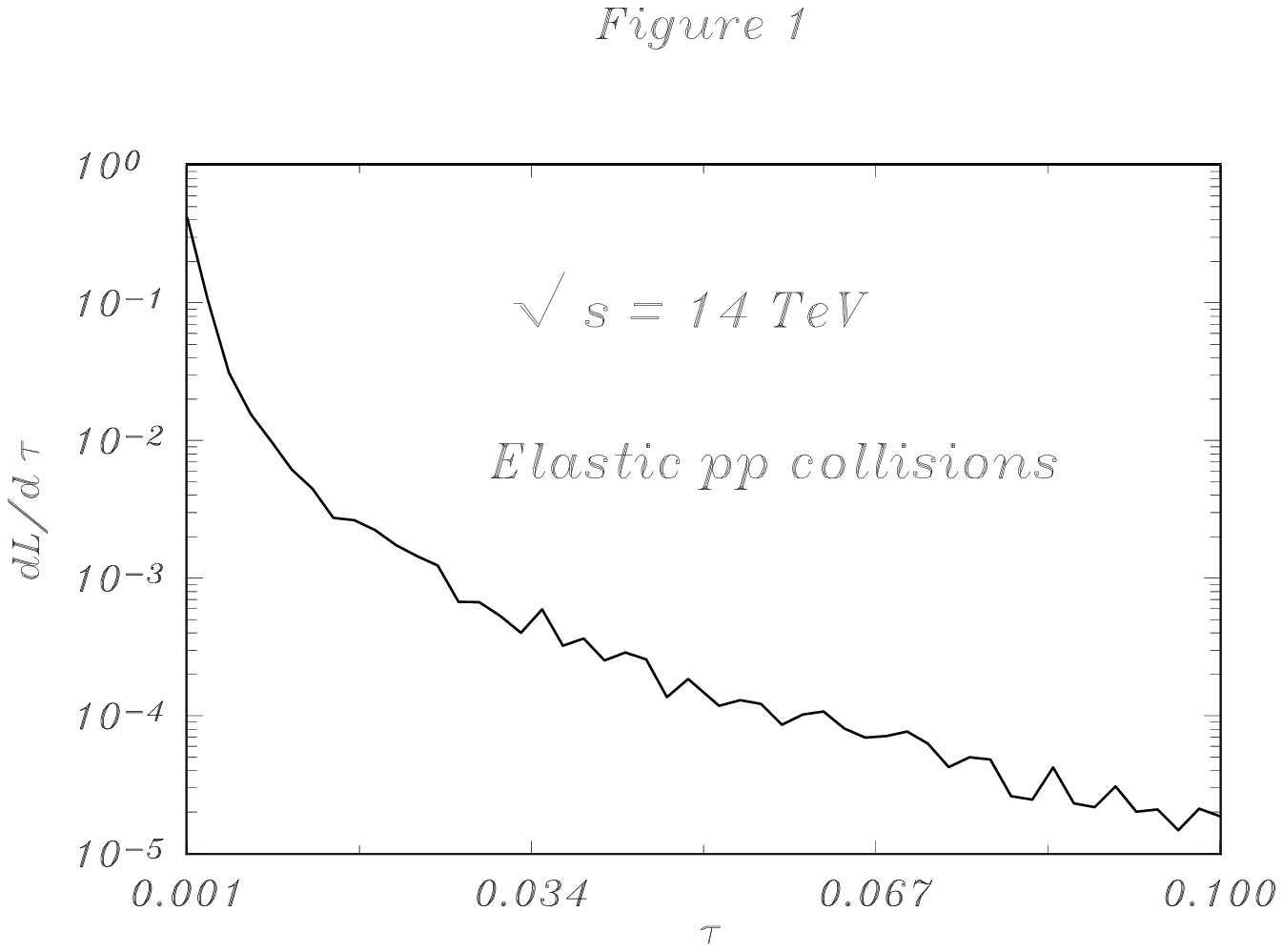}}}

\vspace{1cm}

\end{figure}

\newpage

\begin{figure}[b]
\epsfysize=18cm 
{\centerline{\epsfbox{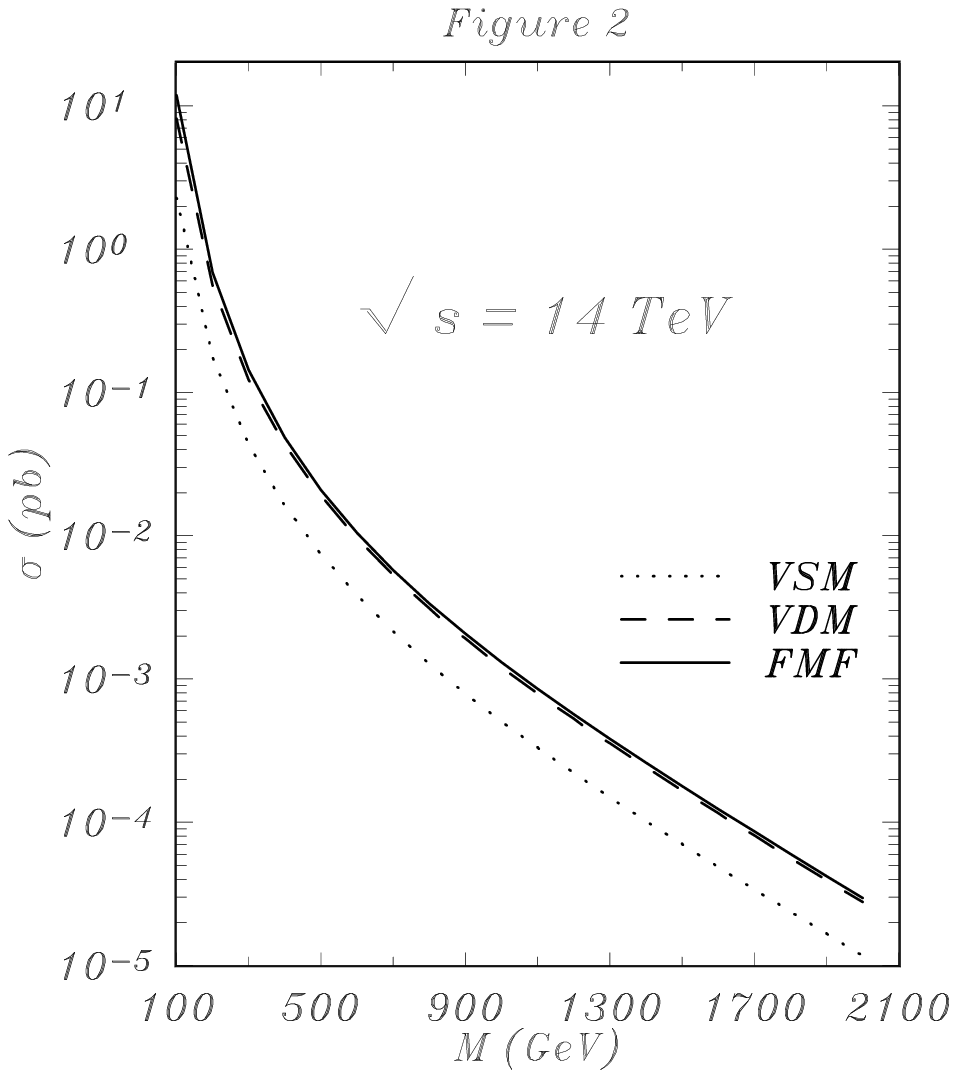}}}

\vspace{1cm}

\end{figure}

\newpage

\begin{figure}[b]
\epsfysize=18cm 
{\centerline{\epsfbox{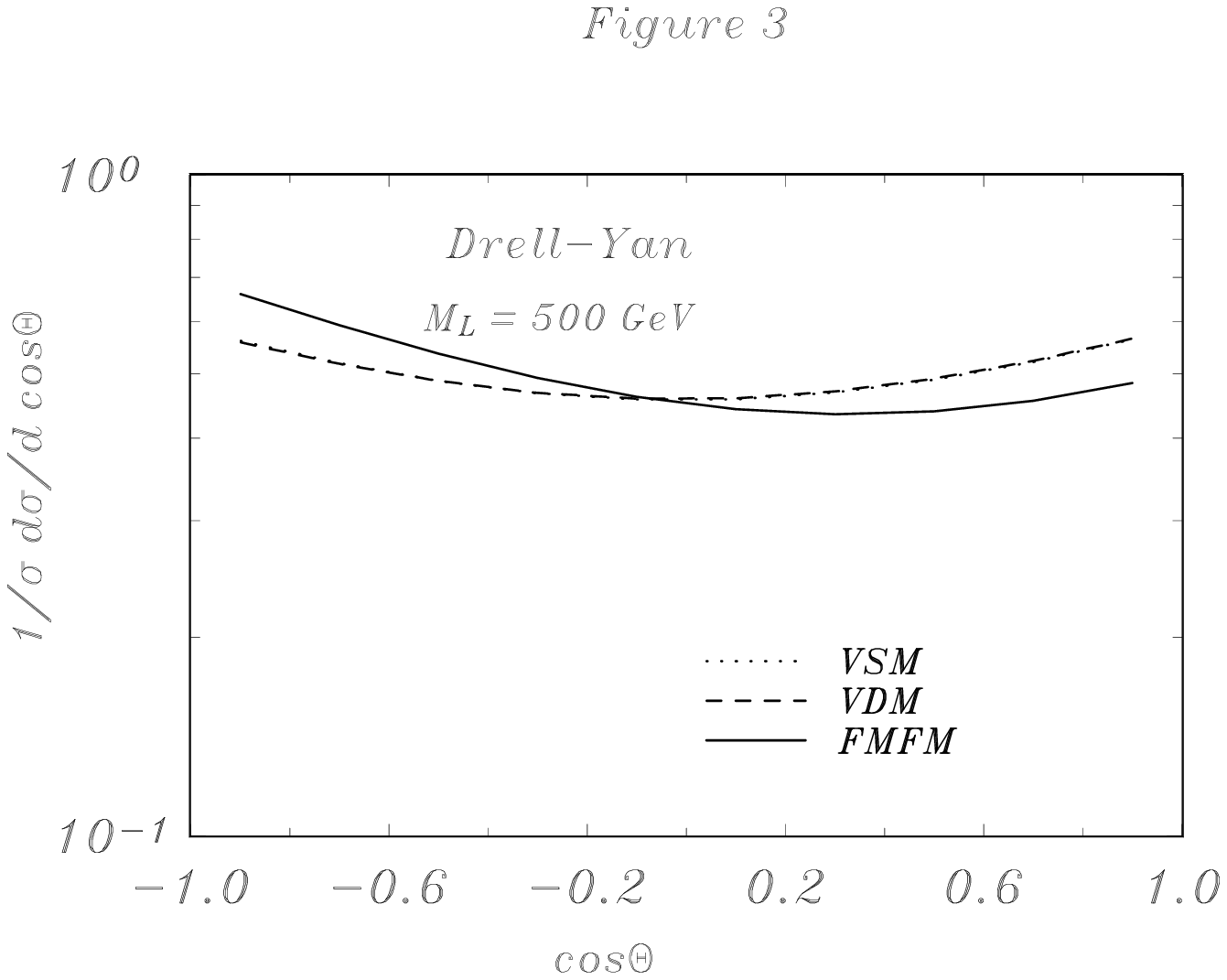}}}

\vspace{1cm}

\end{figure}

\newpage

\begin{figure}[b]
\epsfysize=18cm 
{\centerline{\epsfbox{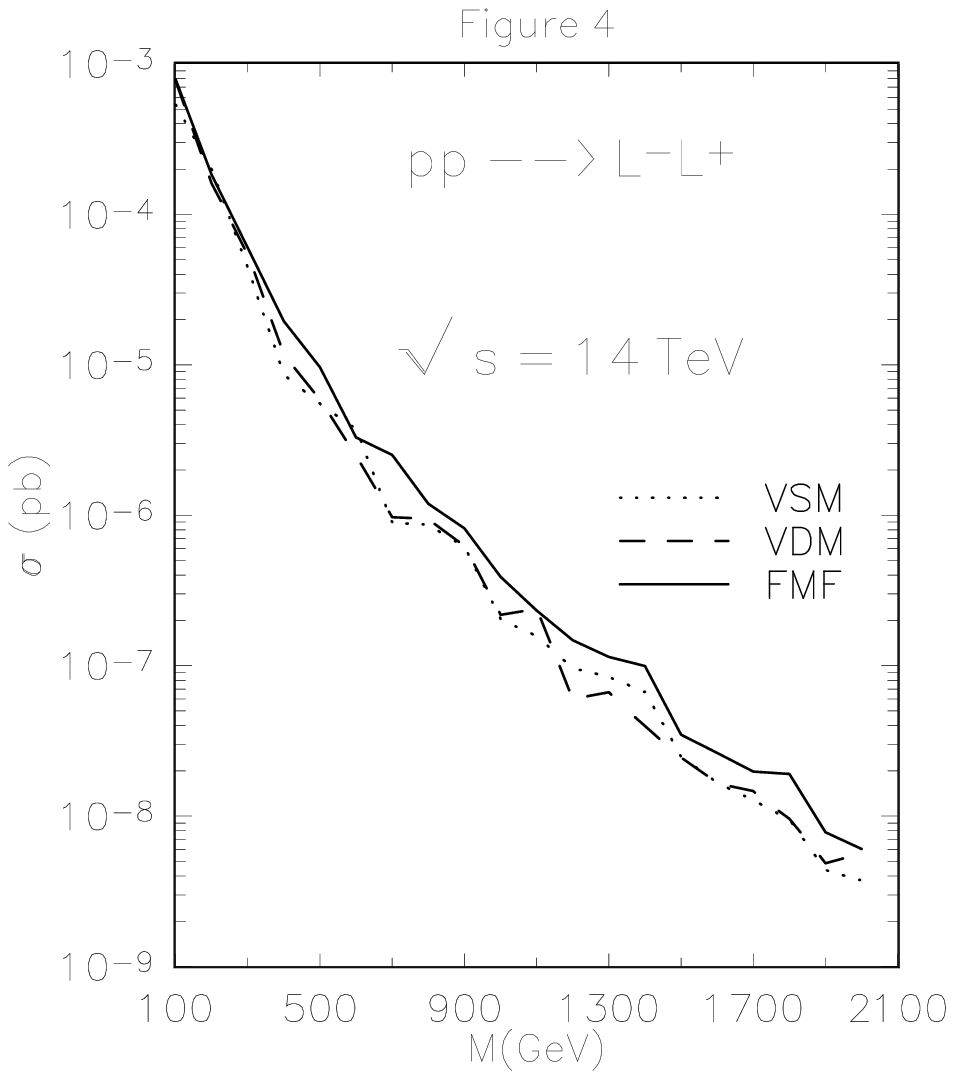}}}

\vspace{1cm}

\end{figure}

\newpage

\begin{figure}[b]
\epsfysize=18cm 
{\centerline{\epsfbox{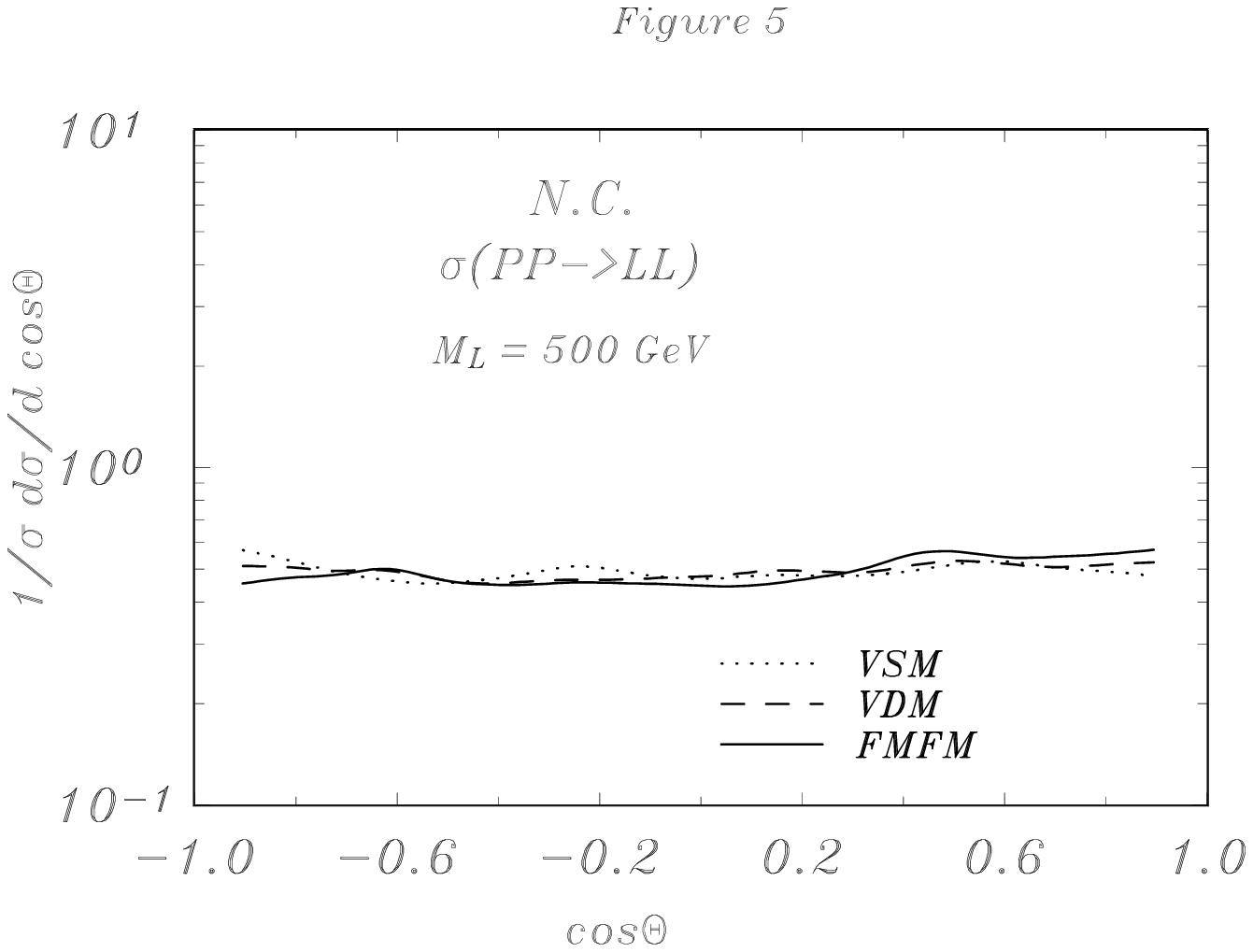}}}

\vspace{1cm}

\end{figure}

\newpage

\begin{figure}[b]
\epsfysize=18cm 
{\centerline{\epsfbox{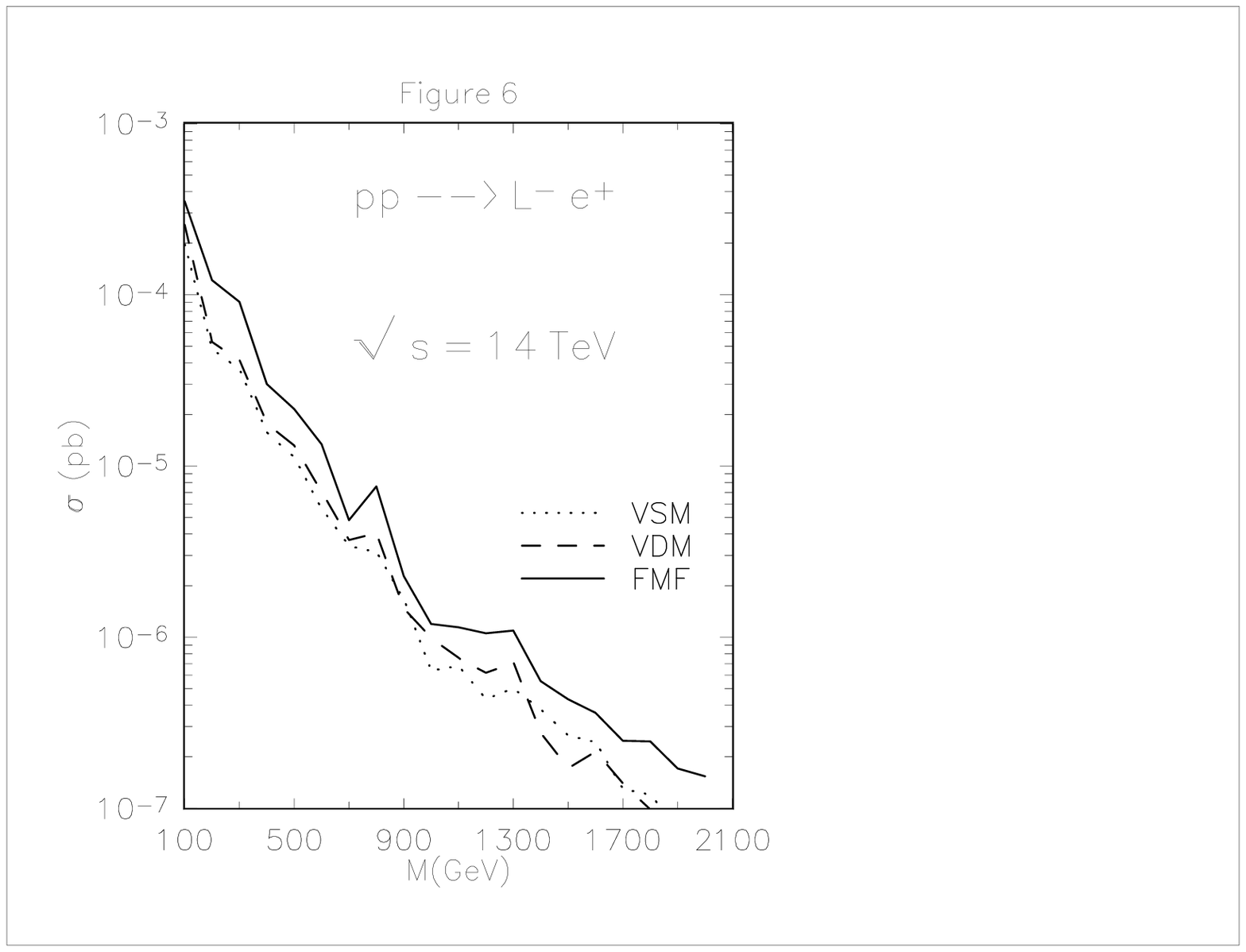}}}

\vspace{1cm}

\end{figure}

\newpage

\begin{figure}[b]
\epsfysize=18cm 
{\centerline{\epsfbox{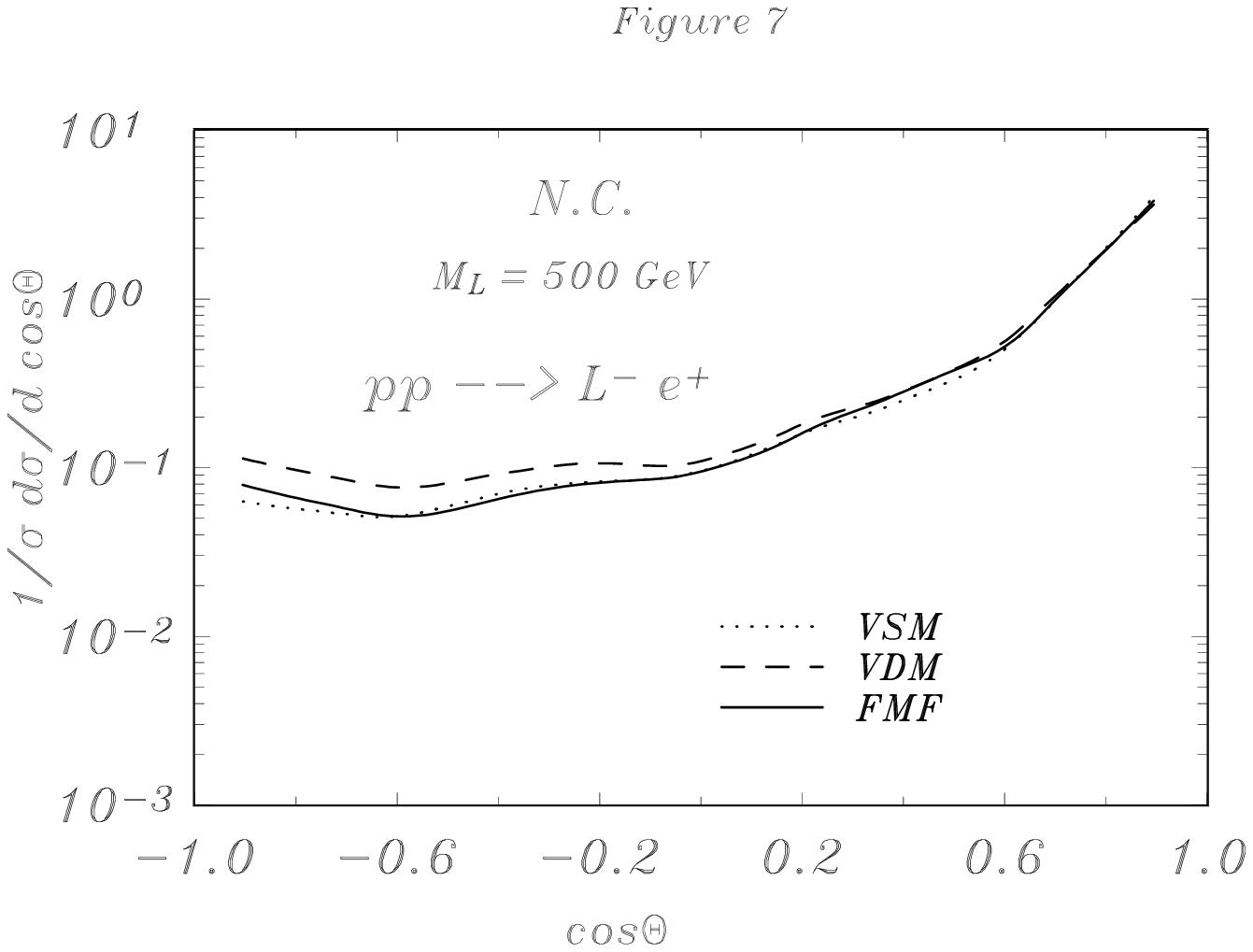}}}

\vspace{1cm}

\end{figure}

\newpage

\begin{figure}[b]
\epsfysize=18cm 
{\centerline{\epsfbox{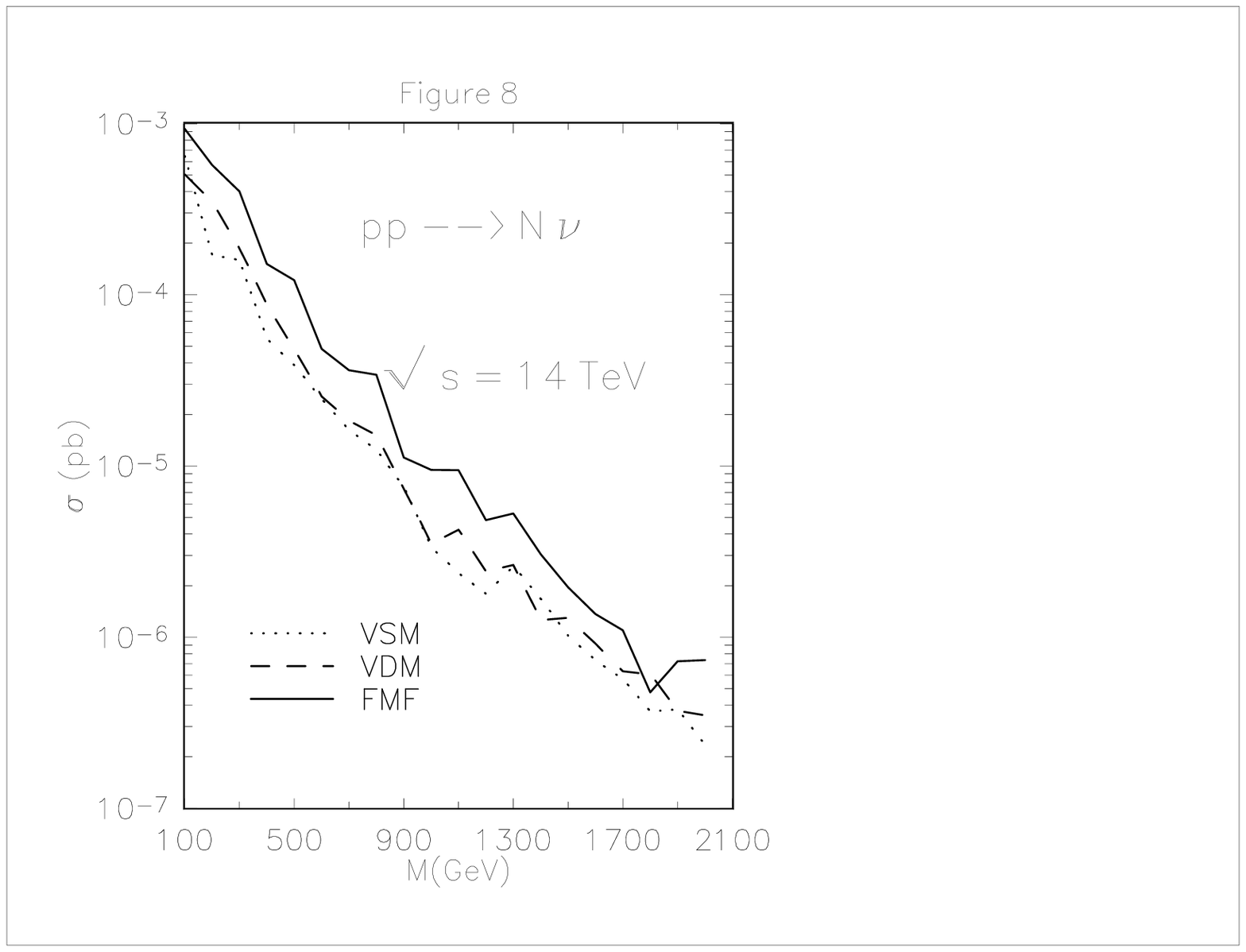}}}

\vspace{1cm}

\end{figure}

\newpage

\begin{figure}[b]
\epsfysize=18cm 
{\centerline{\epsfbox{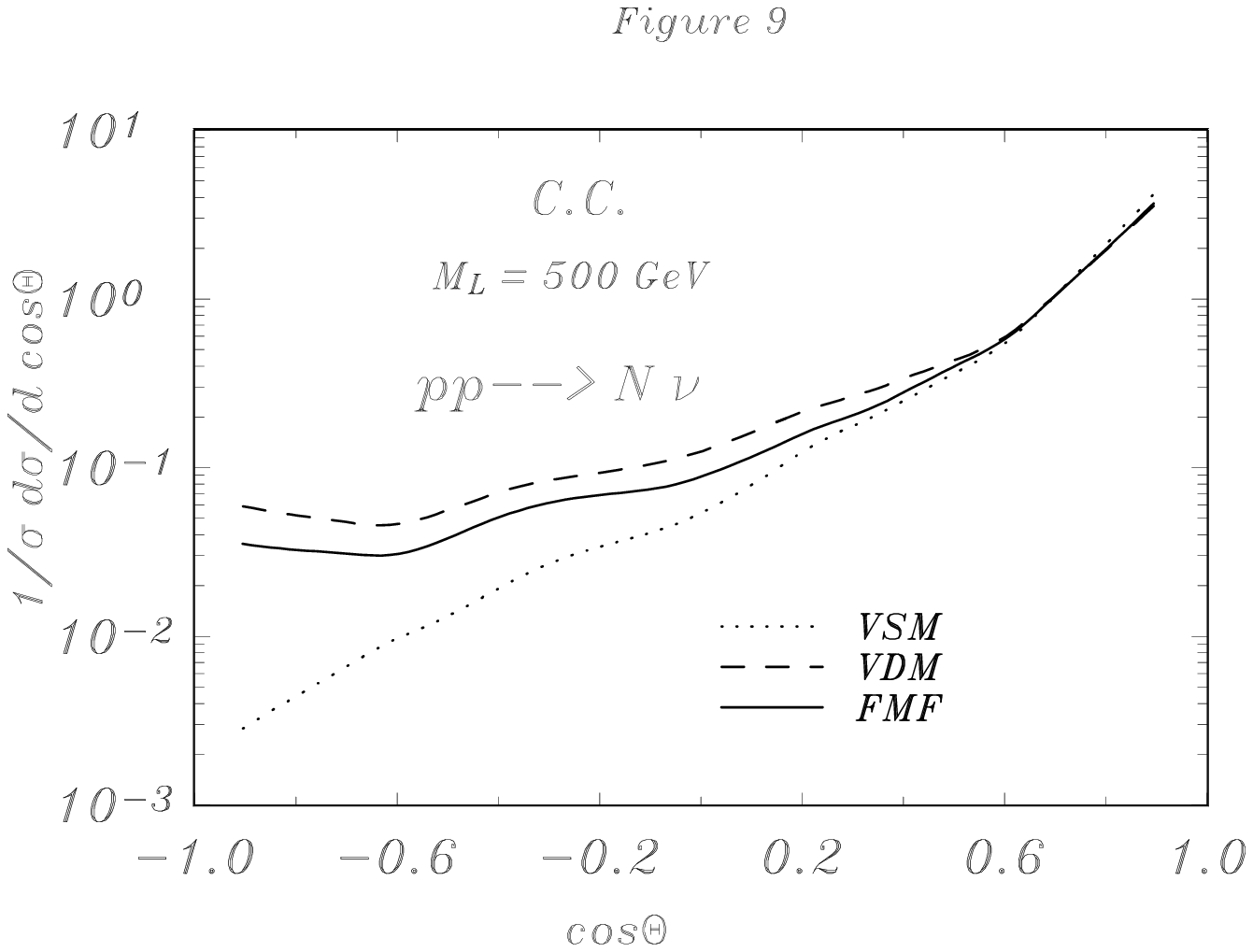}}}

\vspace{1cm}

\end{figure}

\end{document}